\documentclass[a4paper,fleqn]{cas-dc}

\usepackage[numbers]{natbib}
\usepackage{amsmath}
\def\tsc#1{\csdef{#1}{\textsc{\lowercase{#1}}\xspace}}
\tsc{WGM}
\tsc{QE}
\tsc{EP}
\tsc{PMS}
\tsc{BEC}
\tsc{DE}
\begin{document}
\let\WriteBookmarks\relax
\def\floatpagepagefraction{1}
\def\textpagefraction{.001}

\shorttitle{GUSL Pipeline for Prostate Segmentation}

\shortauthors{Jiaxin Yang et~al.}

\title [mode = title]{GUSL: A Novel and Efficient Machine Learning Model for Prostate Segmentation on MRI}                      
%\tnotemark[1,2]


% \tnotetext[1]{This document is the results of the research
%    project funded by the National Science Foundation.}

% \tnotetext[2]{The second title footnote which is a longer text matter
%    to fill through the whole text width and overflow into
%    another line in the footnotes area of the first page.}

\author[1]{Jiaxin Yang}[orcid=0009-0007-4303-1260]



\ead{yangjiax@usc.edu}

%\ead[url]{www.cvr.cc, cvr@sayahna.org}

%\credit{Conceptualization of this study, Methodology, Software}

\affiliation[1]{organization={Ming Hsieh Department of Electrical and Computer Engineering, University of Southern California (USC)},
    addressline={3740 McClintock Ave.}, 
    city={Los Angeles},
    % citysep={}, % Uncomment if no comma needed between city and postcode
    postcode={90089}, 
    state={CA},
    country={USA}}

\affiliation[2]{organization={Department of Urology, Keck School of Medicine, University of Southern California (USC)},
    addressline={1975 Zonal Ave.}, 
    city={Los Angeles},
    % citysep={}, % Uncomment if no comma needed between city and postcode
    postcode={90033}, 
    state={CA},
    country={USA}}

\affiliation[3]{organization={Department of Radiology, Keck School of Medicine, University of Southern California (USC)},
    addressline={1975 Zonal Ave.}, 
    city={Los Angeles},
    % citysep={}, % Uncomment if no comma needed between city and postcode
    postcode={90033}, 
    state={CA},
    country={USA}}


\author[1,2]{Vasileios Magoulianitis}[orcid=0009-0005-3907-1003]
\author[1]{Catherine Aurelia Christie Alexander}[]
\author[1]{Jintang Xue}[orcid=0009-0004-3531-8147]
\author[2]{Masatomo Kaneko}[orcid=0000-0002-1205-807X]
\author[2]{Giovanni Cacciamani}[orcid=0000-0002-8892-5539]
\author[1]{Andre Abreu}[orcid=0000-0002-9167-2587]
\author[3,2]{Vinay Duddalwar}[orcid=0000-0002-4808-5715]
\author[1]{C.-C. Jay Kuo}[]
\author[2]{Inderbir S. Gill}[orcid=0000-0002-5113-7846]
\author[1]{Chrysostomos Nikias}[]

\cortext[cor1]{The corresponding author is with the Electrical Engineering Department of the University of Southern California (USC), Los Angeles, USA.}

%\fntext[fn1]{This is the first author footnote. but is common to third
% \fntext[fn2]{Another author footnote, this is a very long footnote and
%   it should be a really long footnote. But this footnote is not yet
%   sufficiently long enough to make two lines of footnote text.}

% \nonumnote{This note has no numbers. In this work we demonstrate $a_b$
%   the formation Y\_1 of a new type of polariton on the interface
%   between a cuprous oxide slab and a polystyrene micro-sphere placed

\begin{abstract}
Prostate and zonal segmentation is a crucial step for clinical diagnosis of prostate cancer (PCa). Computer-aided diagnosis tools for prostate segmentation are based on the deep learning (DL) paradigm. However, deep neural networks are perceived as "black-box" solutions by physicians, thus making them less practical for deployment in the clinical setting. In this paper, we introduce a feed-forward machine learning model, named Green U-shaped Learning (GUSL), suitable for medical image segmentation without backpropagation. GUSL introduces a multi-layer regression scheme for coarse-to-fine segmentation. Its feature extraction is based on a linear model, which enables seamless interpretability during feature extraction. Also, GUSL introduces a mechanism for attention on the prostate boundaries, which is an error-prone region, by employing regression to refine the predictions through residue correction. In addition, a two-step pipeline approach is used to mitigate the class imbalance, an issue inherent in medical imaging problems. After conducting experiments on two publicly available datasets and one private dataset, in both prostate gland and zonal segmentation tasks, GUSL achieves state-of-the-art performance among other DL-based models. Notably, GUSL features a very energy-efficient pipeline, since it has a model size several times smaller and less complexity than the rest of the solutions. In all datasets, GUSL achieved a Dice Similarity Coefficient (DSC) performance greater than $0.9$ for gland segmentation. Considering also its lightweight model size and transparency in feature extraction, it offers a competitive and practical package for medical imaging applications. 

\end{abstract}

% \includegraphics{figs/grabs.pdf}

% \begin{highlights}
% \item We propose 
% \item Transparent feature extraction
% \item Low complexity and model size
% \item FP reduction in Stage-2 using handcrafted radiomics and anomaly predictions
% \end{highlights}

%%%%%%%%%%%%%%%%%%%%%%%%%%%%%%%%%%%%%%%%%%%%%%%%%%%%%%
% Keywords

%%%%%%%%%%%%%%%%%%%%%%%%%%%%%%%%%%%%%%%%%%%%%%%%%%%%%%
\maketitle
\section{Introduction}\label{sec:introduction}

% Importance of prostate segmentation/cancer
Prostate cancer (PCa) is known as the second most common cancer in men and the fifth leading cause of cancer-related deaths~\citep{chhikara2023global}. There were an estimated $268,490$ of cases and $34,500$ deaths in 2023 due to prostate cancer, comprising $29\%$ of all male cancer cases and contributing to $11\%$ of male cancer-related deaths~\citep{siegel2023cancer}. Yet, early diagnosis of prostate cancer is crucial to improve cure rates and reduce the mortality rate~\citep{cani2022development}. 

% Why accurate MRI prostate segmentation is important to clinicians
In diagnosing prostate disease (e.g., PCa, prostatitis, benign hyperplasia), one of the initial steps taken by the urologists is the calculation of the prostate volume. It is also essential for calculating the PSA density (PSA-D), an important biomarker for screening for clinically significant PCa~\citep{kundu2007prostate}. Prostate segmentation requires manual delineation from physicians on medical images, such as Transrectal Ultrasound (TRUS) or MRI. Undoubtedly, MRI provides significantly better quality images than TRUS and hence it is more useful for PCa diagnosis and measuring prostate's anatomy. For this reason, MRI/TRUS fusion guided-biopsy has been adopted in the clinic, as it provides better accuracy to urologists than TRUS itself~\citep{cool2011fusion}. As such, MRI has become a very useful modality for diagnosing prostate diseases. On the other hand, prostate segmentation is very important for lesions detection, especially if combined with the zonal segmentation for having a more complete clinical diagnosis.

%Also, mpMRI incurs much lower diagnosis costs, as well as less patient discomfort, since TRUS biopsies can have side effects (e.g. infection)~\citep{loeb2013systematic}.

Zonal segmentation highlights the different areas of the prostate and is also important in PCa diagnosis and treatment planning. Statistically, $70\%$ to $80\%$ of prostate cancer cases begin in the peripheral zone (PZ), while only about $20\%$ to $25\%$ arise in the transition zone (TZ)~\citep{ali2022prostate}. Therefore, zonal segmentation plays a significant role to provide clinical priors to urologists~\citep{vargas2012normal}. Notably, the diagnostic visual criteria for TZ and PZ in MRI vary according to the Prostate Imaging Reporting and Data System (PI-RADS). Therefore, it matters for physicians to know which zone they look into. 

% Why automating the prostate segmentation is important and challenging
In clinical practice, physicians manually delineate the prostate on MRI images, which can be time-consuming and labor-intensive, due to the presence of multiple slices and sequences~\citep{steenbergen2015prostate}. In this task, several challenges are posed, including considerable variations in the shape and size of the prostate gland among individuals, unclear boundaries, MRI quality/resolution variations due to different scanner vendors or artifacts due to patient movement ~\citep{ghose2012survey}. Moreover, this process is based on subtle visual features and is highly subjective; thus, the MRI readout may vary according to the experience or training of the physician. Hence, it exhibits significant inter-observer variation~\citep{montagne2021challenge}, which may lead to inconsistencies in calculations. 

% Brief literature review
Early prostate image segmentation algorithms have shown several limitations in practice. They often utilize edge detection filters, thresholding methods, and traditional machine learning techniques. Some of their drawbacks include time-consuming processing, poor generalization, and limited robustness~\citep{juneja2021survey, sharma2024survey}. With the rapid advancement of deep learning (DL), convolutional neural networks (CNNs) have become essential in semantic segmentation, particularly with the U-Net~\citep{unet} model. Diverse variants of U-Net have been developed for medical image segmentation due to their multi-scale feature extraction capabilities~\citep{sun2020saunet, isensee2018nnu, chen2024medical, sultana2020evolution}. DL-based automatic segmentation of prostate and cancer lesions in MRI images has yielded remarkable results, benefiting from the robust hierarchical feature representation. Building on this foundation, researchers have conducted extensive studies to address challenges, such as unclear boundaries at the apex and base of the prostate, along with significant variations in shape and texture among different patients~\citep{zhu2017deeply}. Attention mechanisms~\citep{oktay2018attention, ding2023multi} have also been deployed to allow models to focus on the relevant parts of prostate, thus enhancing the segmentation accuracy.

% Existing prostate segmentation algorithms can be classified into two classes: traditional and deep learning methods. The traditional methods mainly include contour-based \citep{salimi2018fully, ding2003prostate}, atlas-based \citep{klein2008automatic,litjens2012multi,dowling2011fast}, deformable models \citep{cootes1993use,knoll1999outlining}, and machine learning-based models such as c-means clustering \citep{rundo2017automated, rundo2018fully} and classification \citep{allen2006differential, litjens2012pattern}.

% What is the current limitation
Despite their good performance, DL-based methods have certain limitations in healthcare applications. They are often viewed as "black boxes" from clinicians~\citep{adadi2020explainable}, which makes it difficult to understand how they arrive at their decisions. This opacity can be problematic in high-stakes decisions-- which is the case of medical applications-- where interpretability is vital for the adoption of AI-powered tools in the clinical setting. Moreover, a common issue in several medical imaging tasks is the lack of sufficiently large annotated datasets and the class imbalance issue, which is inherent in the medical imaging problem. Besides, DL models entail a high number of learnable parameters and hence a high computational budget. As a consequence, training and inference are expensive in terms of computational resources and energy efficiency~\citep{strubell2020energy}.  

% Introduce Green Learning
Aiming to provide a transparent and computationally efficient machine learning pipeline for prostate segmentation, we introduce GUSL, a feed-forward 3D U-shaped model based on the Green Learning (GL) paradigm~\citep{kuo2022green}. This is a novel representation learning framework based on a linear multi-scale feature extraction model for image analysis. GUSL is an all-regression model, used to predict the residue correction across the different scales in a bottom-up way, from the coarsest to the finer (i.e. original) scale. Its key benefits include the lightweight model size, as well as the transparent feature extraction process. Also, its feed-forward design yields robust features, even when data are scarce. Although the current paper is focused on prostate segmentation, GUSL offers a generic methodology for medical imaging segmentation tasks, which can be seamlessly applied to segment other organs or other medical images (e.g. CT or X-ray). Moreover, to address the class imbalance problem, GUSL is used in a cascaded two-stage approach to mitigate this issue. Our GUSL can achieve a high performance compared with traditional machine learning methods while maintaining transparency in feature extraction and low complexity compared to other DL-based models.

% Contribution
The main contributions can be summarized as follows:
\begin{enumerate}
\item We propose GUSL, a novel machine learning model for 3D medical image segmentation that uses no back propagation or neural networks, but learns the visual representations in a feed-forward manner.

\item A two-stage scheme is devised to address the class imbalance problem and focus the training on the challenging parts of the prostate, such as its boundaries. 

\item A novel multi-level residual correction mechanism is introduced within GUSL to refine the segmentation performance in a coarse-to-fine manner. 

\item Extensive benchmarks are carried out in three different datasets, compared to other DL-based approaches for the gland and zonal segmentation tasks.

\item The proposed GUSL model is lightweight and significantly reduces the model size and complexity compared to other models.
\end{enumerate}

%%%%%%%%%%%%%%%%%%%%%%%%%%%%%%%%%%%%%%%%%%%%%%%%%%%%%%
\section{Related Work}\label{sec:related_work}
%% 1 page

\subsection{Traditional methods}
Prior to the advent of deep learning, the methodologies for prostate segmentation focused on atlas-based approaches, deformable models, and optimization techniques using graph cuts. Atlas-based methods \citep{klein2008automatic, litjens2012multi, dowling2011fast} improve the segmentation accuracy by measuring the similarity between the target image and multiple atlases. \citep{klein2008automatic} generates two deformed atlas images for comparison, emphasizing the significance of the selected atlas images. To measure similarity, they utilize normalized mutual information (NMI) and a majority voting algorithm to combine various image segmentations.

For methods using graph cuts (GC) ~\citep{mahapatra2013prostate}, optimization is used to find the best solution for distinguishing between different regions. Images are modeled as graphs, and segmentation is approached as a graph-cut optimization process constrained by specific conditions. For example, \citep{chen2012medical} combines graph cuts with active appearance models to improve the segmentation accuracy. Another work \citep{qiu2014dual} uses convex optimization to solve the dual problem, employing flow-maximization algorithms in graphs. \citep{tian2017supervoxel} proposes using supervoxel-based graph cuts along with a 3D active contour model to refine segmentation.

\subsection{Learning based methods}
As the amount of annotated MRI data increases, learning-based medical image segmentation methods have made remarkable progress, due to the rapid development of deep learning. Initially, a fully convolutional network (FCN) was proposed by modifying an existing classification CNN for the purpose of segmentation. Several FCN-based methods for medical image segmentation are then proposed. For example, U-Net~\citep{unet} and V-Net~\citep{milletari2016v} are two representative works for 2D and 3D medical image segmentation. U-Net-- which provides a baseline for numerous works-- combines a contracting path of multiple convolutional layers with an expansive path of up-convolutional layers, forming an encoder-decoder network structure. The skip connection architecture in U-Net employs a simple concatenation operation, creating a bridge between the encoder and decoder that leverages both coarse and fine scale features. Due to their robustness, many methods utilize them as backbone models. For example, \citep{sun2020saunet} proposes a shape attentive U-Net (SAUNet) where both the texture information and shape information are used to learn the segmentation. \citep{zhou2019unet++} introduced additional convolutional layers on the skip pathways, intended to bridge the semantic gap between the encoders' and decoders' feature maps. \citep{jin20213d} improved V-Net as 3D PBV-Net with bicubic interpolation. \citep{li2023attention} proposed a network that is based on a double-branch architecture utilizing attention-driven multi-scale learning. Traditional convolutional neural networks (CNNs) are based on linear neuron models, which restrict their capability to capture the complex dynamics of biological neural systems. ~\citep{sumon2025multiclass} combines EfficientNetB4 encoders with decoders based on a Self-organized Operational Neural Network (Self-ONN) to address this limitation by incorporating nonlinear and adaptive operations at the neuron level. 

Attention mechanisms have been shown to be highly effective in learning improved feature representations. \citep{oktay2018attention} proposed attention gates (AGs) to filter the features propagated through the skip connections, which can highlight the important features during learning. \citep{ding2023multi} integrates U-Net skip connections with a multi-scale self-attention mechanism to recalibrate feature maps across multiple layers. \citep{wang2023two} developed a two-stage approach, employing a Squeeze and Excitation (SE) CNN to detect the presence of the prostate in stage one, followed by a Residual-Attention U-Net in stage two to segment the slices that include the prostate gland. \citep{li2023dual} propose a dual attention mechanism using 3D convolutions to learn both gland and lesion segmentation tasks in an end-to-end manner. An interesting work of \citep{jia20193d} introduces 3D APA-Net, a 3D adversarial pyramid anisotropic convolutional deep neural network for prostate segmentation. Moreover, other works~\citep{hung2022cat, chen2023semi} aim to learn the cross-slice features at multiple scales using transformer blocks.

% Boundary
Since prostate segmentation requires accuracy on the boundaries delineation, specific modules are suggested to clarify and define the boundary more precisely. For example, \citep{jia2019hd} proposes a hybrid discriminative network called HD-Net, which has a decoder with two branches: a 3D segmentation branch and a 2D boundary branch. This structure enhances the shared encoder's ability to learn features with greater semantic discrimination. 

%Also, \citep{zhu2019boundary} has proposed BOWDA-Net model, where a boundary-weighted segmentation loss was introduced to the transfer learning.

\subsection{Green Learning methodology}

% SSL methodology
Inspired by the deep learning concept, a new feed-forward machine learning framework without backpropagation, named Green Learning (GL), has been proposed by~\citep{kuo2023green}. In GL-based methods, feature representations are learned in an unsupervised feed-forward manner instead of using backpropagation, utilizing a multi-stage principal component analysis (PCA) for successive subspace learning. It is based on the Saab transform~\citep{kuo2019interpretable}, which is a linear transform. This linearity property enables seamless feature interpretability. Also, this framework has been designed to provide image analysis and feature extraction with fewer parameters and less complexity. Therefore, aiming at providing a more interpretable solution, suitable for healthcare applications and also maintaining the overall complexity low, our proposed work builds upon the GL-framework to inherit its featured advantages.

Transparency and interpretability are of high importance in healthcare tasks, and therefore, a few works have already been published, employing the GL paradigm. \citep{liu2023successive} applies the Saab transform on feature extraction from cine MRI for the task of cardiac disease classification. Also, another work deploys the GL framework on structural MRI for Amyotrophic Lateral Sclerosis (ALS) classification~\citep{liu2021voxelhop}. For prostate in particular, a work for lesion segmentation was initially proposed~\citep{magoulianitis2024pca}. It adopts a two-scale feature extraction model that extracts both spectral and spatial features from the two scales. Regarding prostate gland segmentation, \citep{yang2024pshop} adopts GL and proposes the PSHop model. Our proposed model borrows elements from the latter approach and further proposes new techniques to address challenges, such as class imbalance and accurate prostate boundary delineation.

%%%%%%%%%%%%%%%%%%%%%%%%
\section{Methods}\label{sec:methods}
%% 2.5 pages

\subsection{Two-Stage Cascaded Pipeline}

\begin{figure*}[t]
\begin{center}
\includegraphics[width=1.0\linewidth]{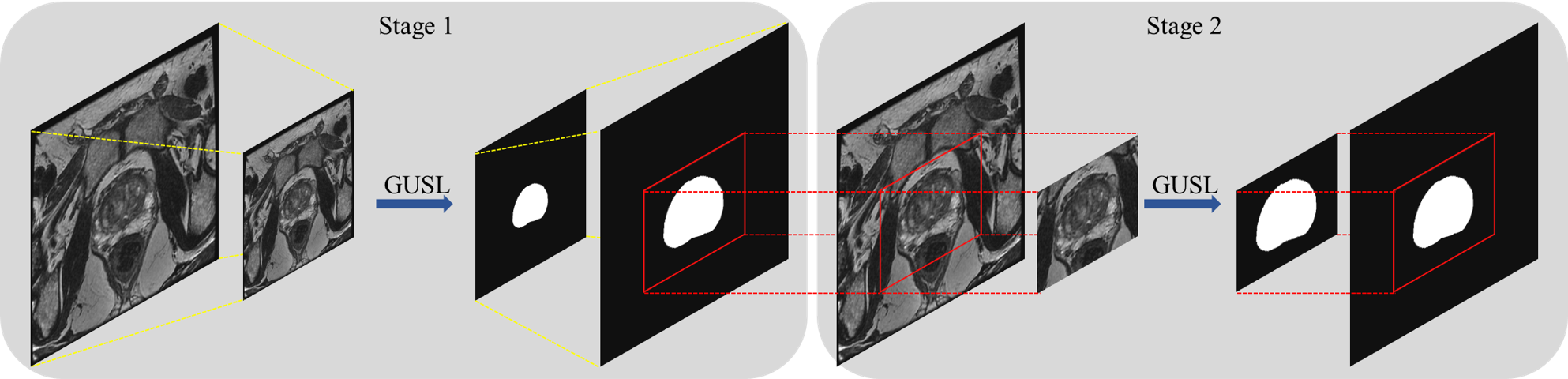}
\end{center}
\caption{Cascading model in two stages. \textbf{Stage-1 (left)}: GUSL segmentation is applied on the downsampled version of the image, and the predicted segmentation is upsampled to the original size. \textbf{Stage-2 (right)}: GUSL segmentation is applied on the resized cropped image, and then concatenated back to the original mask.}
\label{fig:overall_pipeline}
\end{figure*}

% short intro
We propose a two-stage deployment of the GUSL model, as shown in Figure~\ref{fig:overall_pipeline}. In stage 1, we resize the original images to a lower resolution, as shown in Table~\ref{tab:stage_size}. In this way, the rough position and shape of the prostate gland can be determined --which do not need a very detailed information-- thus reducing the overall complexity. The segmentation results of the prostate gland in stage 1 are then upsampled to the original size, and the prostate region is cropped as the Region of Interest (ROI). Based on the rough prediction results, we center crop the prostate. In stage 2, this cropped image will be resized to the same size as shown in Table~\ref{tab:stage_size}. In stage 2, we perform the final segmentation of the prostate gland, as well as the TZ and PZ. Our cascading model is inspired by the U-Net cascade from nnU-Net~\citep{isensee2018nnu}. In stage 1, we use a lower resolution to save some computational complexity for parts of the image that are irrelevant to the prostate and are removed from processing in the next stage. In stage 2, the higher resolution enables a more accurate segmentation and focuses on learning about the difficult parts of the prostate gland. This two-stage approach can balance better the voxel distribution in the second stage, by focusing more on the prostate gland's boundaries, which is an error prone area.

%%%%%%%%%%%%%%%%%%%%%%%%
\begin{table}[t]
\centering
\caption{Size of downsampled images in stage one and size of cropped patches in stage two for different datasets.}
\label{tab:stage_size}
\begin{tabular}{lcc}
\toprule
& Stage 1 & Stage 2 \\
\midrule
T2-cube of Keck & 128x128x64 & 96x96x80 \\
T2-weighted of Keck & 128x128x24 & 128x128x18 \\ 
ISBI-2013 & 128x128x32 & 160x160x16 \\ 
PROMISE12 & 128x128x32 & 160x160x16 \\ 
\bottomrule
\end{tabular}
\end{table}
%%%%%%%%%%%%%%%%%%%%%%%%

%%%%%%%%%%%%%%%%%%%%%%%%%%%%%%%%%%%%%%%%%%%%%%%%%%%%%%
%%%%%%%%%%%%%%%%%%%%%%%%%%%%%%%%%%%%%%%%%%%%%%%%%%%%%%%%%%

%%%%%%%%%%%%%%%%%%%%%%%%%%%%%%%%%%%%%%%%%%%%%%%%%%%%%%%%%%

\subsection{GUSL Pipeline}

\begin{figure*}[t]
\begin{center}
\includegraphics[width=0.9\linewidth]{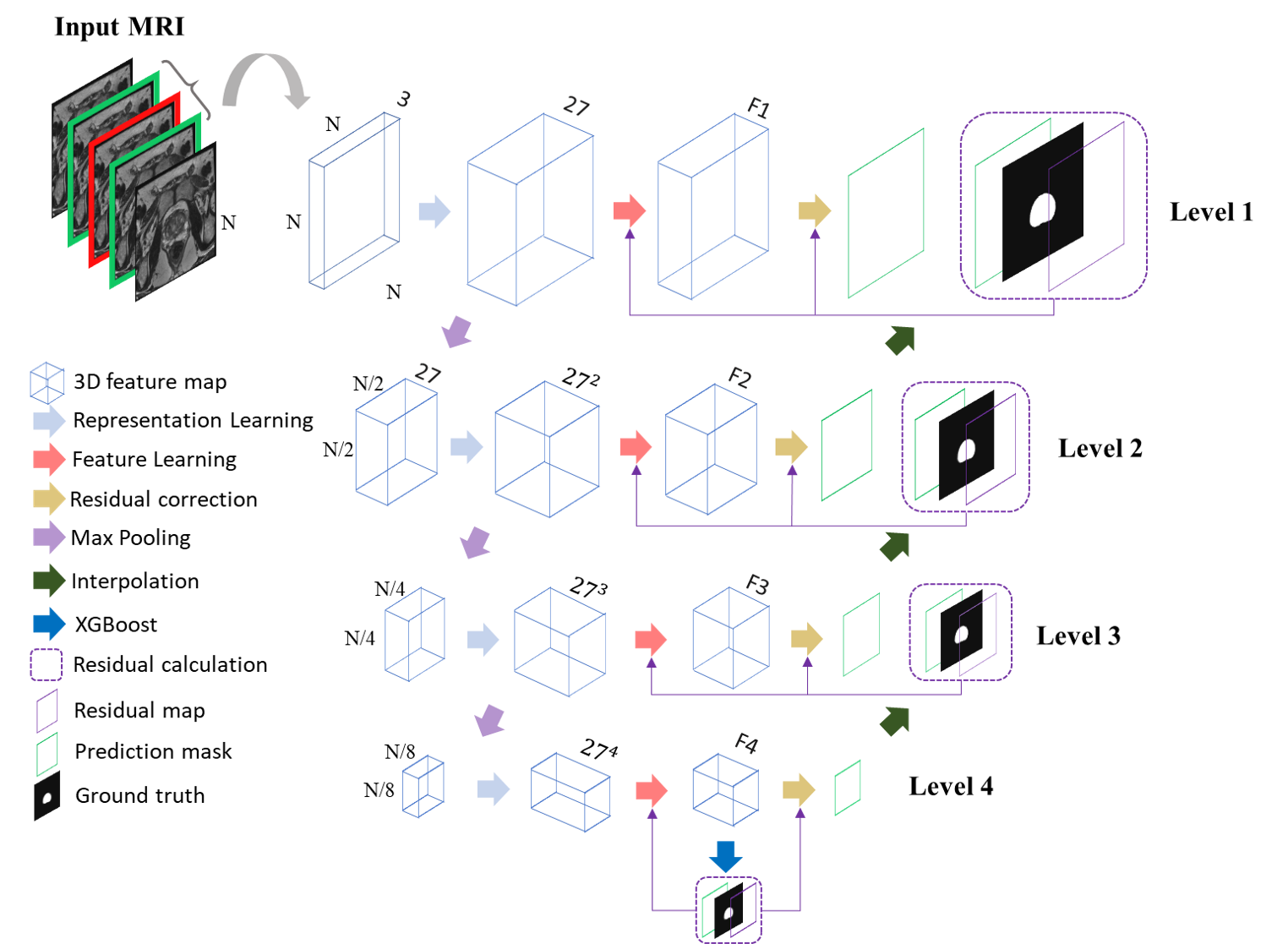}
\end{center}
\caption{Illustration of the 3D architecture of GUSL. Four scales (levels) are used in GUSL. Deeper layers correspond to coarse resolution, while shallower ones to finer resolution. The process demonstrates the
segmentation for one slice of the MRI sequence.}
\label{fig:GUSL_pipeline}
\end{figure*}

Our proposed GUSL method is used in both stages of segmentation, and its architecture overview is shown in Figure~\ref{fig:GUSL_pipeline}, which has multiple levels of different scales, interleaved by max-pooling layers, in the downsampling (encoder) branch and interpolation layers in the upsampling (decoder) branch. The main components at each level of the encoder are: (1) the representation learning (feature extraction) by using the channel-wise Saab transform, (2) feature learning to enhance the feature space using the least-squares normal transform, and the relevant feature test for filtering out noisy or irrelevant features for the target task. In the decoder's side, we perform residual correction to improve the predicted mask from the previous level in a recursive way. In particular, at level 4, an XGBoost regressor was used to generate an initial low-resolution prediction mask. During the residual calculation, the difference between the prediction and ground truth masks results in the residual map, which can be regarded as the regression target. To create regression targets for each level, we simply downsample the ground truth mask using patch averaging to create a continuous range of the voxel targets. The details of each module mentioned are described in the following subsections.
%%%%%%%%%%%%%%%%%%%%%%%%%%%%%%%%%%%%%%%%%%%%%%%%%%%%%%%%%%

%%%%%%%%%%%%%%%%%%%%%%%%%%%%%%%%%%%%%%%%%%%%%%%%%%%%%%
%%%%%%%%%%%%%%%%%%%%%%%%%%%%%%%%%%%%%%%%%%%%%%%%%%%%%%%%%%

%%%%%%%%%%%%%%%%%%%%%%%%%%%%%%%%%%%%%%%%%%%%%%%%%%%%%%%%%%
\subsubsection{Unsupervised Representation Learning}\label{subsec:representation learning}
We consider the segmentation task as a voxel-wise regression problem and extract the representation of neighboring voxels (window) at different scales. Unlike deep learning-based methods, which learn convolutional filters through end-to-end optimization of the loss function, we utilize cascaded VoxelHop units~\citep{liu2021voxelhop}, which is a statistical approach to extract feature vectors in a feed-forward manner. The representation learning process operates entirely without supervision, requiring no labels during training. 

\begin{figure*}[t]
\begin{center}
\includegraphics[width=0.8\linewidth]{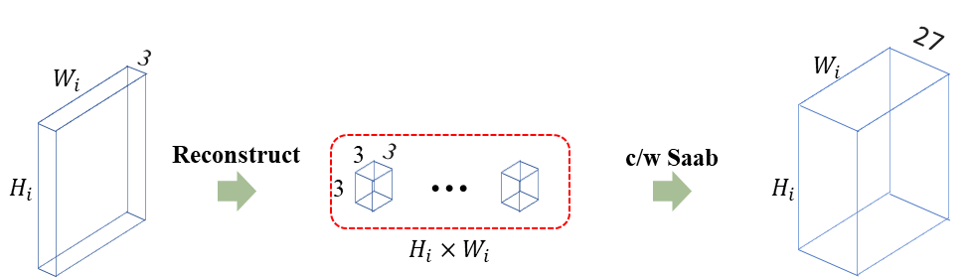}
\end{center}
\vspace{-3mm}
\caption{Neighborhood reconstruction into cuboid blocks for representation learning and feature extraction in GUSL.}
\label{fig:voxelhop}
\end{figure*}

% \subsubsubsection{Feature Extraction with VoxelHop}
Figure~\ref{fig:voxelhop} illustrates the process of each VoxelHop unit. It involves two consecutive steps: 1) neighborhood reconstruction in 3D space, and 2) learning feature representations using the c/w Saab transform~\citep{chen2020pixelhop++}. 

Suppose the input tensor of the $i$-th VoxelHop unit has dimension $H_i\times W_i\times C_i \times K_i$, where $H_i$, $W_i$, and $C_i$ represent the resolution in 3D space, and $K_i$ represents the dimension of the feature vector for each voxel extracted from the
$(i-1)$-th VoxelHop unit. Specifically, for the first VoxelHop unit where $i=1$, $H_1\times W_1\times C_1$ corresponds to the input MRI data resolution, and $K_1=1$. At first, after padding along the z-axis, the input tensor will be divided into $H_i\times W_i\times 3 \times K_i$ blocks. After padding along the X and Y axes, we construct the neighborhood in the 3D block centered about each voxel, which turns out in $Hi\times Wi$ blocks. The neighborhood size is fixed as $3\times 3\times 3$ in spatial. That is, each voxel in the neighborhood has a feature vector of dimension $K_i$, which results in a tensor of size $H_i\times W_i\times (27\times K_i)$. The filter sizes and strides of each layers are included in Table~\ref{tab:table_voxelhop}.

%%%%%%%%%%%%%%%%%%%%%%%%%%%%%%%%%%%%%%%%%%%%%%%%%%%%%%
%%%%%%%%%%%%%%%%%%%%%%%%%%%%%%%%%%%%%%%%%%%%%%%%%%%%%%%%%%

%%%%%%%%%%%%%%%%%%%%%%%%%%%%%%%
\begin{table}[t]
\centering
\caption{Parameters in VoxelHop for each layer.}
\label{tab:table_voxelhop}
\begin{tabular}{llll}
\toprule
  & Filter Size &  Stride \\
\midrule
VoxelHop 1        & $(3\times3)\times3$&  $(1\times1)\times1$& \\
Max-pooling 1     & $(2\times2)\times1$&  $(2\times2)\times1$& \\
VoxelHop 2        & $(3\times3)\times3$&  $(1\times1)\times1$& \\
Max-pooling 2     & $(2\times2)\times1$&  $(2\times2)\times1$& \\
VoxelHop 3        & $(3\times3)\times3$&  $(1\times1)\times1$& \\
Max-pooling 3     & $(2\times2)\times1$&  $(2\times2)\times1$& \\
VoxelHop 4        & $(3\times3)\times3$&  $(1\times1)\times1$& \\ 
\bottomrule
\end{tabular}
\end{table}
%%%%%%%%%%%%%%%%%%%%%%%%%%%%%%%

% \subsubsubsection{Channel-wise Saab Transform}
The channel-wise Saab transform~\citep{chen2020pixelhop++} is an essential component of the VoxelHop unit. It is a data-driven transform based on the principal component analysis (PCA). It transforms an input tensor from the spatial to the spectral domain. In the channel-wise version of the transform, the Saab transform is applied independently for each $C_i$ channel. Suppose the input vector is $\mathbf{x}\epsilon \mathbb{R}^{N}$, where $N=3\times 3 \times 3$. We can extract features by projecting the input vector onto the anchor vectors learned from PCA, expressed as an affine transformation:
\begin{equation}\label{eq:affine_transform}
y_m = \mathbf{a}_m^T\cdot \mathbf{x} + b_m, m=0,1,\cdots, M-1,
\end{equation}
where $\mathbf{a}_m$ is the $m$-th anchor vector of dimension $N$, and $M$ is the total number of anchor vectors. In this context, the channel-wise Saab transform is a data-driven method for learning anchor vectors from the formed tensors, gathered from the input data. After neighborhood reconstruction and forming the 3D input blocks, we separate the input subspaces into a direct sum of two distinct subspaces, labeled DC and AC, as shown in Eq.~\ref{eq:SDC_SAC}. The terminology is derived from "direct circuit" and "alternating circuit" concepts in circuit theory.
\begin{equation}\label{eq:SDC_SAC}
    \hspace{20mm}   S = S_{DC} \oplus S_{AC}
\end{equation}
$S_{DC}$ and $S_{AC}$ are spanned by $DC$ and $AC$ anchor vectors, defined as:
\begin{itemize}
\item DC anchor vector $\mathbf{a}_0=\frac{1}{\sqrt{N}}\left ( 1, 1, \cdots , 1 \right )^T$
\item AC anchor vectors $\mathbf{a}_m$, $m=1,\cdots, M-1$
\end{itemize}

The input signal $\mathbf{x}$ is projected onto $\mathbf{a}_0$ to get the DC component $\mathbf{x}_{DC}$. Then the AC component is extracted by subtracting the DC component from the input signal, i.e. $\mathbf{x}_{AC}=\mathbf{x}-\mathbf{x}_{DC}$. 

The AC anchor vectors are learned by performing principal component analysis (PCA) on the AC component. The first $K$ principal components are retained as the AC anchor vectors. Thus, features could be extracted by projecting $\mathbf{x}$ onto the above learned anchor vectors based on Eq.~\ref{eq:affine_transform}. Since PCA yields orthogonal anchor vectors, the extracted features will be independent and hence uncorrelated. Finally, the bias term is selected to ensure all features are positive by following~\citep{kuo2019interpretable}.

%%%%%%%%%%%%%%%%%%%%%%%%%%%%%%%%%%%%%%%%%%%%%%%%%%%%%%
% \begin{figure}[t]
% \begin{center}
% \includegraphics[width=1.0\linewidth]{figures/PCA_Saab.pdf}
% \end{center}
% \caption{An illustration of the local neighborhood construction for unsupervised filter learning using the Saab transform based on PCA. Each anchor vector corresponds to a subspace on the N-dimensional plane (denoted in different colors). The input image is projected on to these subspaces to obtain its spectral decomposition at a certain scale that corresponds to the output feature map. In c/w Saab this decomposition is applied on every single feature map in a recursive manner, until the maximum number of layers is reached.}
% \label{fig:voxelhop_unit}
% \end{figure}
%%%%%%%%%%%%%%%%%%%%%%%%%%%%%%%%%%%%%%%%%%%%%%%%%%%%%%%%%%

\subsubsection{Receptive Field Expansion}\label{subsec:expansion}
\begin{figure*}[t]
\begin{center}
\includegraphics[width=1.0\linewidth]{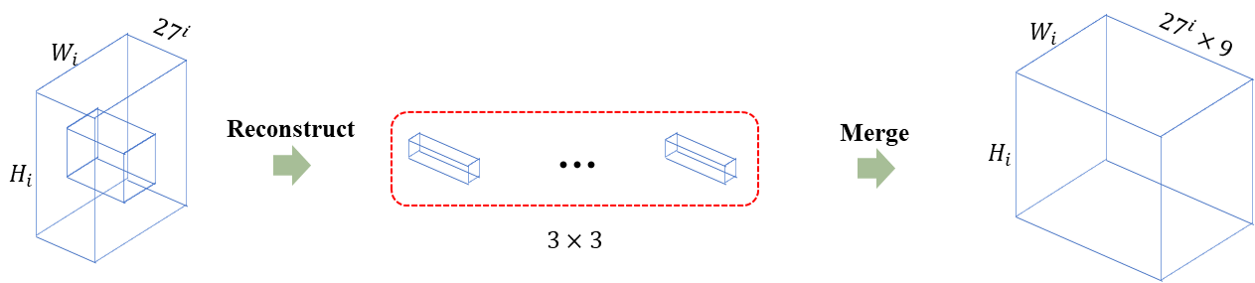}
\end{center}
\vspace{-4mm}
\caption{Receptive field expansion improves the number of features for each voxel from $\mathbf27^i$ to $\mathbf27^i\times9$ and provides more context to the regressor.}
\label{fig:merge}
\end{figure*}

Providing more context about each voxel, definitely helps in the segmentation task. In each layer, we merge features from a surrounding $3\times3$ region into the center voxel. That is, we extract features from the $3\times3$ voxel region, and flatten them to form nine $1$D vectors. By concatenating them, we grow the feature representation of the center voxel ($\times 9$), expanding also the receptive field about it, thus providing more context that can help discriminate over the difficult textures between background and foreground (Figure~\ref{fig:merge}). 

%%%%%%%%%%%%%%%%%%%%%%%%%%%%%%%%%%%%%%%%%%%%%%%%%%%%%%
%%%%%%%%%%%%%%%%%%%%%%%%%%%%%%%%%%%%%%%%%%%%%%%%%%%%%%%%%%

\subsubsection{Supervised Feature Learning}\label{subsec:feature learning}

In GL-based representation learning, the features is the spatial-spectral representation of the input image and hence it is derived in an unsupervised way. Yet, some of these features may not be robust for the specific task, because of the lack of supervision. To enhance the VoxelHop's feature space, a supervised feature learning module is introduced, where the features from VoxelHop provide a basis to generate new and more robust features by using the Least-squared Normal Transform (LNT)~\citep{wang2023enhancing}, which introduces supervision in the generation of new features. Features from VoxelHop and LNT are concatenated, and then a Relevant Feature Test (RFT)~\citep{yang2022supervised} module is used to filter out the dimensions that are less informative. Both LNT and RFT are supervised processes, where labels are used to make the feature representation of each voxel more compact and discriminant for the specific task.

% \subsubsection{Least-squares Normal Transform (LNT)}
Least-squares Normal Transform (LNT)~\citep{wang2023enhancing} utilizes linear transformation to generate new features. Firstly, all VoxelHop features are divided into K feature subsets by XGBoost~\citep{chen2016xgboost}. Then, a linear regression model is trained on each subset independently. For the features in each subset, we find the corresponding weights from the linear regression and utilize them for a linear combination, which will generate one new feature. Hence, for K subsets, we can generate K new features, based on the original VoxelHop features. The new LNT linear projections of the original features are more discriminant, since they are optimized for a certain task. In our experiment, we set $K=400$.

\begin{figure*}[t]
\begin{center}
\includegraphics[width=0.9\linewidth]{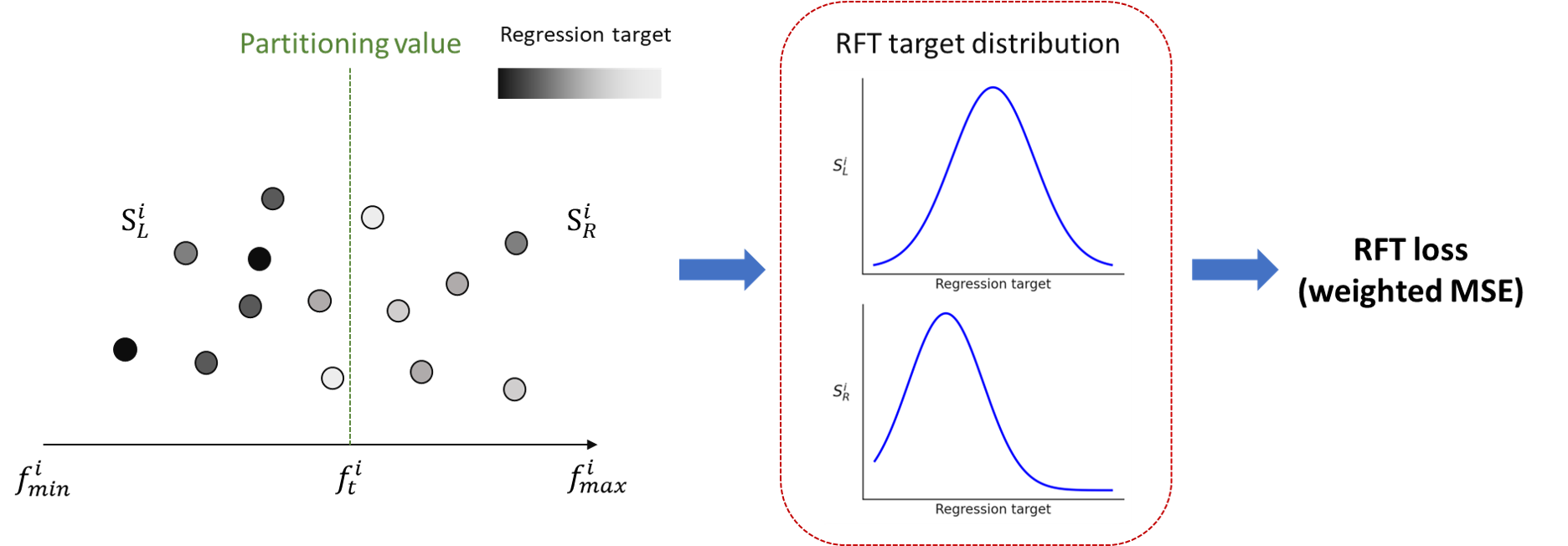}
\end{center}
\caption{An overview of the Relevant Feature Selection (RFT) method. For the i-th feature, RFT measures the weighted estimated regression MSE in both sets as the RFT loss.}
\label{fig:rft}
\end{figure*}

% \subsubsection{Relevant Feature Test (RFT)}
As shown in Figure~\ref{fig:rft}, Relevant Feature Test (RFT)~\citep{yang2022supervised} could help us select discriminant features from both VoxelHop and LNT features, thus filtering out dimensions that are noisy and increase unnecessarily the complexity. For regression tasks, mapping the input features to a target scalar function is more efficient when the feature dimensions can separate samples into segments with lower variances. This is because the regressor can use the average of each segment as the target value from $f^{i}_{min}$ to $f^{i}_{max}$, for the $i$-th feature, while its corresponding variance indicates the mean-squared error (MSE) of the segment. Driven by this observation and the binary decision tree, RFT divides a feature dimension into left and right segments ($S^i_L$ and $S^i_R$) by partitioning value $f^i_t$ and assesses the total MSE from both. We use this approximation error as the RFT loss function (weighted MSE). The smaller the RFT loss, the more discriminant the feature dimension is. RFT is used in GUSL to reduce the overall complexity and provide a more compact and discriminant feature representation.

%%%%%%%%%%%%%%%%%%%%%%%%%%%%%%%%%%%%%%%%%%%%%%%%%%%%%%
%%%%%%%%%%%%%%%%%%%%%%%%%%%%%%%%%%%%%%%%%%%%%%%%%%%%%%%%%%

\subsubsection{Regression \& Residual Correction}\label{subsec:residual correction}

\begin{figure*}[t]
\begin{center}
\includegraphics[width=0.7\linewidth]{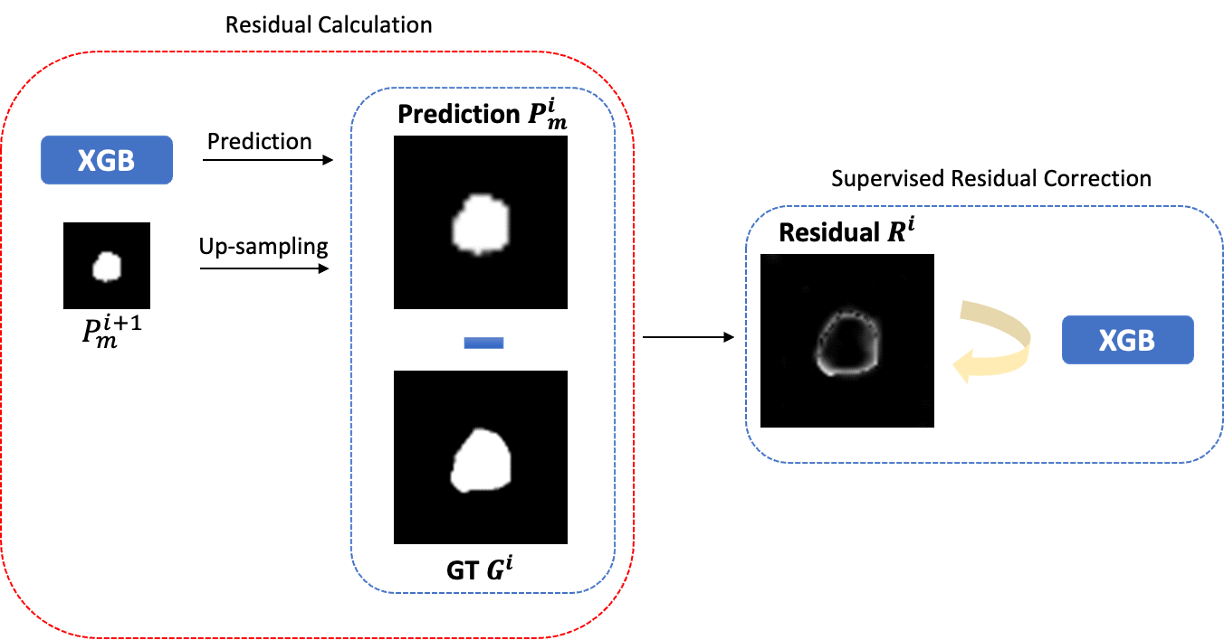}
\end{center}
\vspace{-4mm}
\caption{The pipeline for residual correction in the $i^{th}$ level.}
\label{fig:rc}
\end{figure*}

Having obtained an optimal feature representation for each voxel, an important part of the GUSL model is the residue prediction at each layer in a bottom-up way, to predict the final segmentation result. We contend that viewing the problem as regression --instead of classification-- helps to control the segmentation errors throughout the scales and refine the predictions in the finer resolution layers. As shown in Figure~\ref {fig:rc}, in the bottom level, an initial XBGoost regressor~\citep{chen2016xgboost} is used to obtain an initial prediction mask $P^i_m$ at the coarsest resolution level $i=4$. This XGB regressor makes the first prediction of the prostate at the coarsest scale (Fig.~\ref{fig:GUSL_pipeline} - blue arrow). The label $G^i$ is a continuous value and obtained by successively downsampling the original ground truth mask using average pooling. The initial XGB model is trained on a more balanced voxel distribution, after center cropping the prostate gland. The targets are per-voxel continuous probability values.

% Motivation about the 2nd XGB correction
An issue that arises in prostate segmentation is that certain areas of the prostate gland are more challenging than others. For example, the boundaries of the prostate have a larger variation across different slices, and also they are statistically less significant, compared to the pure background and foreground patches. Therefore, it is desired to control which areas of the prostate require more attention, when they yield a higher error rate. 

By using a second regression model (Fig.~\ref{fig:GUSL_pipeline} - yellow arrows) in the same layer to correct the error prone regions, we can provide residual correction on the challenging parts of the image and further improve the $P^i_m$ prediction. The difference between the two XGB regression models is that the initial one is trained with samples from the entire input cropped image, while the second XGB is meant to correct its predictions in the more error prone regions $B^i$, which are usually the boundaries of the prostate. The regression target in the second regression model is the residual map $R^i$ instead of ground truth map $G^i$. In other words, both XGB share the same feature representation, but they are trained on samples from different regions, as well as different targets. This can be viewed as an efficient attention mechanism, where the first XGB makes an initial segmentation of the prostate, while the second XGB is used to further correct the errors of the previous one, which in practice occur on the boundaries as shown in $R^i$. In other words, the basis regression model's predictions are viewed as a prior information for the second model to further correct its errors. Figure~\ref {fig:rc} illustrates an example of a typical case for the second XGB targets.  

To propagate the corrected segmentation to the next layer, the predicted mask $P^i_r$ is interpolated into the resolution to a higher level $i-1$ to $P^{i-1}_m$. In turn, residual correction is conducted at this level to get an improved prediction mask $P^{i-1}_r$. This process is repeated for each subsequent level until level 1. Therefore, the regression models from layers $3$ to $1$ have as a task to compensate for the error between the interpolated segmentation predictions $P^i_m$ and the downsampled ground truth $G^i$ of the layer. Inside each residual correction block, we first try to reduce the number of samples and make the distribution of samples more balanced by ROI selection, which reduces the model size and improves the performance. 

%%%%%%%%%%%%%%%%%%%%%%%%%%%%%%%%%%%%%%%%%%%%%%%%%%%%%%
%%%%%%%%%%%%%%%%%%%%%%%%%%%%%%%%%%%%%%%%%%%%%%%%%%%%%%%%%%

\subsubsection{Region of Interest (ROI) selection}

\begin{figure*}[t]
\begin{center}
\includegraphics[width=0.8\linewidth]{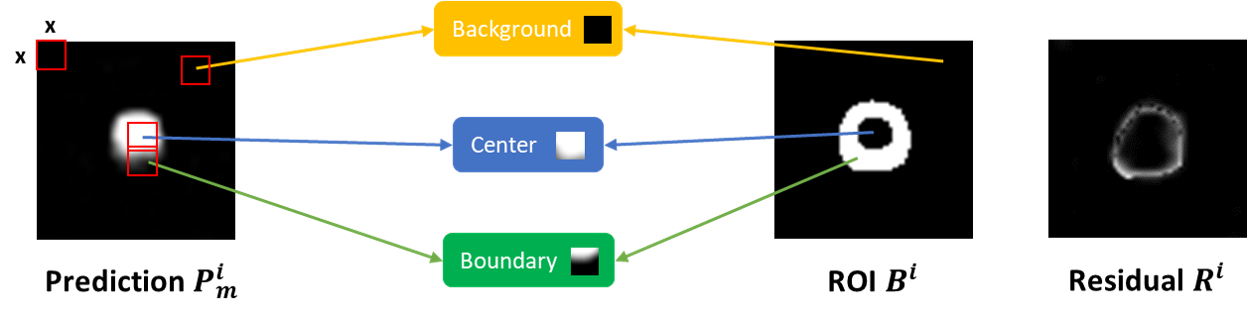}
\end{center}
\vspace{-4mm}
\caption{The process of ROI selection to balance the sample distribution for training.}
\label{fig:roi}
\end{figure*}

One essential step for efficient attention on the prostate boundaries is the ROI selection, so that the error correction focuses on the error-prone areas. In practice, this process samples mostly from the prostate boundary region as the ROI region $B^i$ for level $i$. For the initial prediction $P^i_m$, we divide it into three parts: the background, prostate gland, and boundary. The background and prostate gland voxels are predicted with very small error and hence the energy in the residual map on these areas is almost zero. On the other hand, the block of the boundary is between the background and the prostate is of higher energy, because of the higher error along the boundaries. Therefore, the second XGB model will be trained on patches that mostly belong on the boundary or any other area that requires correction, as it is directed by the energy of error. Figure~\ref {fig:roi} illustrates the sampling over different areas of the image.

%After that, we assign one(white) to the boundary blocks, zero(black) to the other two blocks. In this way, we could obtain binary mask ROI $B^i$. Compared with the residual map $R^i$, the ROI $B^i$ contains all residual regions.

%Since this process is unsupervised, we could find ROI $B^i$ for both the training set and the testing set. For voxels in ROI $B^i$, samples from the white region will be used for training and testing. 

%%%%%%%%%%%%%%%%%%%%%%%%%%%%%%%%%%%%%%%%%%%%%%%%%%%%%%
%%%%%%%%%%%%%%%%%%%%%%%%%%%%%%%%%%%%%%%%%%%%%%%%%%%%%%%%%%

\subsubsection{Error Compensation}

During inference, the residual correction can improve the up-sampling predicted mask across layers. That is, for level $i$, we will have an initial prediction $P^i_m$ from XGBoost regression or up-sampling $P^{i+1}_m$. The prediction mask in level $i$ will be a probability map. We also down-sample the ground truth to the resolution of level $i$, and the down-sampled ground truth is $G^i$, which will also be a probability map including values from zero to one. During training, the residual map $R^i$ is expressed as:
\begin{equation}\label{eq:residual_training}
\hspace{25mm} R^i = G^i - P^i_m
\end{equation}

\begin{figure*}[t]
\begin{center}
\includegraphics[width=1.0\linewidth]{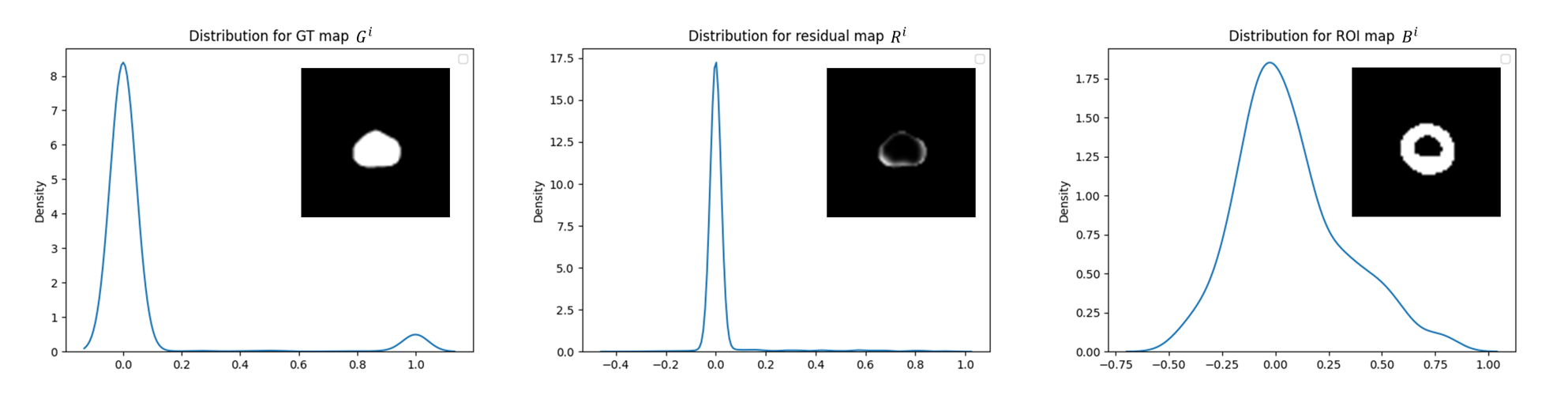}
\end{center}
\vspace{-4mm}
\caption{The values distribution for the ground truth map $G^i$, residual map $R^i$, and ROI map $B^i$.}
\label{fig:distribution}
\end{figure*}

Then, the residual map $R^i$ will be used as a target to train an XGBoost regression for samples from the ROI region $B^i$. For the testing process, we will extract all samples from the ROI region, and get the predicted residual $R^i$ with XGBoost regression. In doing so, as shown in Figure~\ref{fig:distribution}, we can transfer the imbalanced distribution of ground truth map $G^i$, which has two peaks for background and prostate, to the residual map $R^i$, which has a symmetrical Gaussian distribution. However, the center peak for the Gaussian distribution of $R^i$ is still too high. Thus, we use ROI selection to provide attention and focus the training on the boundary region $B^i$. After ROI selection, the distribution of $R^i$ is transferred to a more symmetric Gaussian distribution for ROI map $B^i$. This highlights our motivation in training a second XGB regressor on the residual map. After getting the predicted residual map $R^i$, it is added to the initial prediction $P^i_m$ from the initial XGB model (level-4) or up-sampling $P^{i+1}_m$ and get the improved prediction map $P^i_r$ as 

 \begin{equation}\label{eq:residual_testing}
    \hspace{25mm}   P^i_r = P^i_m + R^i
\end{equation}   

All in all, GUSL is an all-regression model architecture, however the models are used at different targets, such as predicting a coarse segmentation of the gland (basis model), correct errors on the boundaries (level-4 residue correction model) and compensate the errors of the up-sampling interpolated predictions (models in levels 1-3). The model in level-1 yields the final probability map for the prostate gland. After a probability thresholding one obtains the final predicted segmentation. 

For the zonal segmentation task, GUSL is applied in the same way as for the gland segmentation. Two different GUSL models are trained to predict the TZ and PZ masks, having also available their corresponding annotation masks.

%%%%%%%%%%%%%%%%%%%%%%%%%%%%%%%%%%%%%%%%%%%%%%%%%%%%%%
%%%%%%%%%%%%%%%%%%%%%%%%%%%%%%%%%%%%%%%%%%%%%%%%%%%%%%%%%%

\section{Experimental Setup}\label{sec:experiments_setup}

\subsection{Database}
% dataset
To demonstrate the effectiveness of the proposed GUSL method, we conduct experiments based on two public MRI databases, NCI-ISBI 2013 Challenge (Automated Segmentation of Prostate Structures \citep{bloch2015nci}), PROMISE-12 Challenge (Prostate MR Image Segmentation 2012\citep{litjens2014evaluation}), and a private dataset, named USC-Keck, with patient data from the Keck School of Medicine at the University of Southern California. 

The ISBI-2013 dataset consists of 60 cases of axial T2-weighted MR 3D series, where half were obtained at 1.5T (Philips Achieva at Boston Medical Center) and the other half at 3T (Siemens TIM at Radboud University Nijmegen Medical Center). The pixel spacing within each slice ranges from 0.39 \textit{mm} to 0.75 \textit{mm}, while the through-plane resolution ranges from 3.0 \textit{mm} to 4.0 \textit{mm} among different patients. 

The Prostate MR Image Segmentation (PROMISE12) challenge was established to assess both interactive and automatic algorithms for prostate segmentation, focusing on their performance and robustness. The challenge dataset comprises 50 transversal T2-weighted MR images collected from multiple centers, featuring variations in scanner manufacturers, field strengths, and imaging protocols.

The USC-Keck dataset includes a cohort of 600 patients with T2-w, where a subset of 260 patients have also available T2-cube series. Specifically, for T2-w pixel spacing ranges from 0.5 \textit{mm} to 0.7 \textit{mm} and the through-plane resolution (z-axis) from 3.0 mm \textit{mm} to 4.0 \textit{mm}. For T2-Cube, the pixel spacing is set at 0.83 \textit{mm} and the through-plane resolution (z-axis) at 1.4 \textit{mm}. The scanner used to acquire those images was a GE 3T with 8ch Cardiac coil. In our experimental analysis, we provide results for both the T2-Cube and T2-w cohorts within the USC-Keck dataset to offer more inclusive results and analysis.

\subsection{Pre-processing}
% preprocessing
For both datasets, we first normalize the resolution of different images by resampling using Lanczos interpolation onto the same physical resolution, as shown in Table~\ref{tab:stage_size}. Here, the through-plane resolution is increased, so that the segmentation in the 3D space can be more accurate. To reduce the artifacts during image acquisition, contrast enhancement using CLAHE~\citep{CLAHE} was applied.

\subsection{Evaluation Metrics}
% evaluation
We use three different metrics to evaluate the performance for segmentation results. The first one is the Dice Similarity Coefficient (DSC) (see Equation~\ref{eq:dsc}) in a binary scenario, where $X$ and $Y$ represent the predicted segmentation mask and ground truth, respectively. It is commonly used in image segmentation tasks to compare the overlap between the predicted segmentation and the ground truth. It calculates the ratio of the intersection of two binary sets against their average cardinality.

\begin{equation}
    \hspace{14mm}   DSC\left (X,Y\right ) = \frac{2\left | X\cap Y \right |}{\left | X \right|+\left | Y \right |}
    \label{eq:dsc}
\end{equation}

The Average Boundary Distance (ABD)~\citep{heimann2009comparison} (shown in Equation~\ref{eq:abd}) is a metric that assesses the accuracy of boundaries in image segmentation tasks. In its formula, $\partial A$ and $\partial B$ are the sets of boundary points of the predicted segmentation mask and ground truth, and $d(x, \partial Y)$ refers to the shortest Euclidean distance from point $x$ to boundary $\partial Y$. It measures the average distance between the boundary points of the predicted segmentation and those of the ground truth segmentation. ABD quantifies how closely the predicted boundary aligns with the ground truth boundary. Unlike overlap-based metrics, such as the DSC, ABD specifically focuses on contour or edge alignment. This makes it particularly useful for evaluating fine details along boundaries.

\begin{equation}
    ABD = \frac{1}{|\partial A| + |\partial B|} \left( \sum_{x \in \partial A} d(x, \partial B) + \sum_{y \in \partial B} d(y, \partial A) \right)\,
    \label{eq:abd}
\end{equation}

 The third metric used in our validation experiments is the 95\% Hausdorff Distance (HD95)~\citep{huttenlocher1993comparing}. It is a boundary-based metric used to evaluate the similarity between two shapes or segmentations, especially in medical image analysis. One needs first to determine the distance from each point in one set to the nearest point of the other set. The 95th percentile value of these distances, yields the final HD95 value.

\subsection{Experimental Details}
% setup
In all three datasets, we apply the same experimental settings for both gland and zonal segmentation tasks. We apply 5-fold cross validation on the training images and calculate the mean and standard deviation of the evaluation scores. For benchmarking GUSL with other DL-based methods, we conduct the same experiment setting in both datasets and train V-Net and U-Net-based architectures.

Besides the performance evaluation and benchmarking with other DL-based models, we also compare the model size and complexity, in terms of the number of parameters and floating-point operations per second (FLOPS).

%For PROMISE12, since it still receives submission of the test set predictions, we train PSHop using the 50 training images, and evaluate the 10 test results on the challenge website. 

% Besides single observer, we consider the second condition where inter-observer generalizability is evaluated. In the first experiment under this condition, we merge the training images of both datasets into one set with 110 images. Then we apply 3-fold cross validation by following~\citep{jia20193d}. In the second experiment, we train on the 50 training images from PROMISE12, and evaluate on the 5 folds of the 60 images from ISBI-2013. The mean and standard deviation results are then compared with the ISBI-2013 single observer experiment. 

\section{Experimental Results}\label{sec:results}
%% 2 page
\subsection{Quantitative Results}
To compare GUSL's performance against other methods, we examined both baseline and state-of-the-art models in both tasks, and across all metrics and datasets. As such, the average validation performance of DSC, HD95, and ABD, is reported along with the standard deviation, so we can draw more concrete conclusions. Looking into the results, a general observation is that GUSL achieves a state-of-the-art performance in most of the comparisons. We can see that it outperforms by large margins in certain cases-- most of the state-of-the-art DL-based methodologies.

%%%%%%%%%%%%%%%%%%%%%%%%
\begin{table}[t]
\centering
\scriptsize
\caption{Comparison with the DSC metric between GUSL and two popular baseline deep learning models on the T2-cube cohort of the USC-Keck dataset.}
\label{tab:keck_t2cube_dsc}
\begin{tabular}{lccc}
\toprule
& Whole gland & TZ & PZ \\
\midrule
2D U-Net~\citep{unet} & 0.809 $\pm$ 0.036 & 0.741 $\pm$ 0.041 & 0.525 $\pm$ 0.032 \\ 
PSHop~\citep{yang2024pshop} & 0.873 $\pm$ 0.017 & 0.845 $\pm$ 0.025 & 0.656 $\pm$ 0.012 \\ 
V-Net~\citep{milletari2016v} & 0.906 $\pm$ 0.009 & 0.878 $\pm$ 0.019 & 0.747 $\pm$ 0.014 \\
\midrule
\textbf{GUSL} & \textbf{0.9314 $\pm$0.004} & \textbf{0.8964 $\pm$ 0.009} & \textbf{0.759 $\pm$ 0.013} \\ 
\bottomrule
\end{tabular}
\end{table}
%%%%%%%%%%%%%%%%%%%%%%%%

%%%%%%%%%%%%%%%%%%%%%%%%
\begin{table}[t]
\centering
\normalsize
\caption{Comparison with DSC and HD95 metric between GUSL and two popular baseline deep learning models on the T2-w cohort of the USC-Keck dataset.}
\label{tab:keck_t2w}
\begin{tabular}{lcc}
\toprule
& DSC$\uparrow$ & HD95$\downarrow$ \\
\midrule
2D U-Net~\citep{unet} & 0.863 $\pm$ 0.079 & 2.771 $\pm$ 1.972 \\
V-Net~\citep{milletari2016v} & 0.884 $\pm$ 0.078 & 3.979 $\pm$ 6.721 \\
\midrule
\textbf{GUSL} & \textbf{0.891 $\pm$0.070} & \textbf{2.042 $\pm$3.421} \\ 
\bottomrule
\end{tabular}
\end{table}
%%%%%%%%%%%%%%%%%%%%%%%%

\textit{ISBI-2013 dataset~\citep{bloch2015nci}}: The DSC and ABD metrics are used to compare GUSL and several popular deep learning models. The results are available in Table~\ref{tab:isbi_whole} for the gland segmentation, Table~\ref{tab:isbi_tz} for the TZ segmentation, and Table~\ref{tab:isbi_pz} for the PZ segmentation. From the comparisons, for the gland segmentation task, GUSL outperforms all other models in both metrics, achieving the highest DSC and the lowest ABD scores. Regarding the zonal segmentation task, our model achieves the best performance in segmenting the TZ. On the PZ segmentation, GUSL has a very competitive standing among the other models, where the 3D attention U-Net achieves the best performance.   

\textit{PROMISE12 dataset~\citep{litjens2014evaluation}}: The DSC and HD95 metrics were used to compare GUSL with the rest DL-based methods in Table~\ref{tab:PROMISE12}. GUSL also achieves a state-of-the-art performance among most of the other methods. Only nnU-Net achieves a slightly better performance, yet the difference is very small. 

%%%%%%%%%%%%%%%%%%%%%%%%
\begin{table}[t]
\centering
\normalsize
\caption{Comparison with DSC and ABD metric between GUSL and several popular baseline deep learning models~\citep{ocal2022novel} for ISBI-2013 dataset with prostate gland.}
\label{tab:isbi_whole}
\begin{tabular}{lcc}
\toprule
& DSC$\uparrow$ & ABD$\downarrow$ \\
\midrule
V-Net~\citep{milletari2016v} & 0.762 & 3.562 \\
PSHop~\citep{yang2024pshop} & 0.826  & 2.901 \\
3D GCN~\citep{peng2017large} & 0.831 & 2.204 \\
3D-UNet~\citep{unet} & 0.866 & 1.848 \\ 
VoxResNet~\citep{chen2018voxresnet} & 0.89 & 1.375 \\ 
3D APA-Net~\citep{jia20193d} & 0.893 & 1.167 \\ 
\midrule
\textbf{GUSL} & \textbf{0.905} & \textbf{1.119} \\ 
\bottomrule
\end{tabular}
\end{table}
%%%%%%%%%%%%%%%%%%%%%%%%

%%%%%%%%%%%%%%%%%%%%%%%%
\begin{table}[t]
\centering
\caption{Comparison with DSC and ABD metric between GUSL and several popular baseline deep learning models~\citep{qin20203d} for ISBI-2013 dataset with TZ.}
\label{tab:isbi_tz}
\begin{tabular}{lcc}
\toprule
& DSC$\uparrow$ & ABD$\downarrow$  \\
\midrule
PShop~\citep{yang2024pshop} & 0.667 & 3.210 \\
V-Net~\citep{milletari2016v} & 0.796 & 4.051 \\
3D U-Net~\citep{unet} & 0.827 & 2.517 \\ 
3D R2U-Net~\citep{alom2019recurrent} & 0.833 & 2.205 \\
3D Attention U-Net~\citep{schlemper2019attention} & 0.860 & 1.903 \\
\midrule
\textbf{GUSL} & \textbf{0.875} & \textbf{1.210} \\ 
\bottomrule
\end{tabular}
\end{table}
%%%%%%%%%%%%%%%%%%%%%%%%

\textit{USC Keck dataset}: The results for the private  USC-Keck dataset using DSC and HD95 metrics are shown in Tables ~\ref{tab:keck_t2cube_dsc} and \ref{tab:keck_t2w} for the T2-Cube and T2-w cohorts, respectively. GUSL outperforms all three other models in both gland and zonal segmentation tasks. Notably, in the T2-Cube cohort, GUSL surpasses the performance of 2D U-Net and the V-Net by large margins, especially in the gland segmentation. Another observation made is that GUSL has a more stable performance across the different training/validation folds, as inferred by the lower standard deviation.

%%%%%%%%%%%%%%%%%%%%%%%%
\begin{table}[t]
\centering
\caption{Comparison with DSC and ABD metric between GUSL and several popular baseline deep learning models~\citep{qin20203d} for ISBI-2013 dataset with PZ.}
\label{tab:isbi_pz}
\begin{tabular}{lcc}
\toprule
& DSC$\uparrow$ & ABD$\downarrow$ \\
\midrule
PShop~\citep{yang2024pshop} & 0.366 & 3.226 \\
V-Net~\citep{milletari2016v} & 0.741 & 5.332 \\
3D R2U-Net~\citep{alom2019recurrent} & 0.768 & 4.802 \\
3D U-Net~\citep{unet} & 0.775 & 4.756 \\ 
3D Attention U-Net~\citep{schlemper2019attention} & \textbf{0.790} & 4.085 \\
\midrule
\textbf{GUSL} & 0.774 & \textbf{1.734} \\ 
\bottomrule
\end{tabular}
\end{table}
%%%%%%%%%%%%%%%%%%%%%%%%

%%%%%%%%%%%%%%%%%%%%%%%%
\begin{table}[t]
\centering
\normalsize
\caption{Comparison with DSC and HD95 metric between GUSL and several popular baseline deep learning models~\citep{bhandary2023investigation} for PROMISE12 dataset.}
\label{tab:PROMISE12}
\begin{tabular}{lccc}
\toprule
& DSC$\uparrow$ & HD95$\downarrow$ \\
\midrule
V-Net~\citep{milletari2016v} & 0.865 $\pm$ 0.142 & 2.844 $\pm$ 0.921 \\
3D-UNet~\citep{unet} & 0.865 $\pm$ 0.141 & 2.861 $\pm$ 0.830 \\ 
U-Net++~\citep{zhou2019unet++} & 0.878 $\pm$ 0.100 & 2.809 $\pm$ 0.825 \\
SegResNet~\citep{myronenko20183d} & 0.893 $\pm$ 0.085 & 1.573 $\pm$ 0.909 \\ 
Attention U-Net~\citep{oktay2018attention} & 0.900 $\pm$ 0.091 & \textbf{1.436 $\pm$ 0.740} \\ 
Z-Net~\citep{zhang2019prostate} & 0.905 $\pm$ 0.030 & 4.410 $\pm$ 2.000 \\ 
nnU-Net~\citep{isensee2018nnu} & \textbf{0.910 $\pm$ 0.121} & 9.735 $\pm$ 7.566 \\
\midrule
\textbf{GUSL} & 0.907 $\pm$0.024 & 1.607 $\pm$0.804 \\ 
\bottomrule
\end{tabular}
\end{table}
%%%%%%%%%%%%%%%%%%%%%%%%

\subsection{Qualitative comparison}

%%%%%%%%%%%%%%%%%%%%%%%%
\begin{figure*}[t]
\begin{center}
\includegraphics[width=1.0\linewidth]{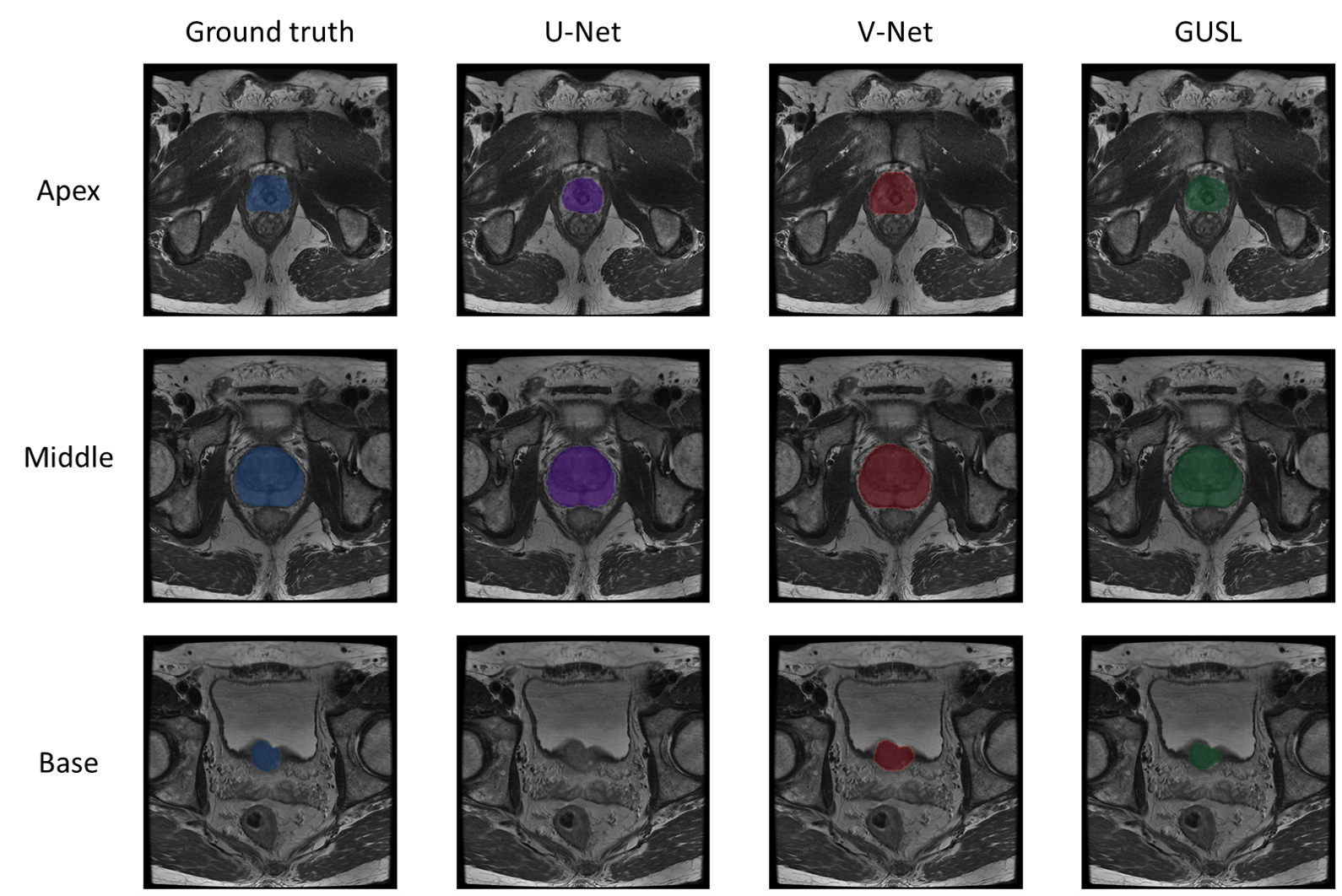}
\caption{Qualitative comparison between GUSL and other DL methods on the whole gland segmentation task for T2-weighted MRIs in the USC-Keck dataset.} \label{fig:keck_t2w}
\end{center}
\end{figure*}

Performance comparisons are not limited to the quantitative performance benchmarking, but we also visually compare the output of GUSL against other baseline DL-based models, to provide a more comprehensive comparison.
Some exampled from the prostate gland segmentation task are illustrated in Figure~\ref{fig:keck_t2w}. We have chosen to include examples of different parts of the prostate through the z-plane, such as the apex, middle, and base. This is because the appearance varies significantly across the in-plane slices and thus provides different challenges to the segmentation module. One observation is that most models achieve a satisfactory performance in the apex and middle areas, whereas they are more challenged in the base area of the prostate. In particular, U-Net misses the entire prostate region, V-Net shows over-segmentation effects, while GUSL achieves a segmentation closer to the ground truth mask, with some under-segmentation effects. Segmenting these areas is more challenging due to their complex shapes,  smaller target regions, cluttered appearance, and more variations across different patients.

\begin{figure*}[t]
\begin{center}
\includegraphics[width=1.0\linewidth]{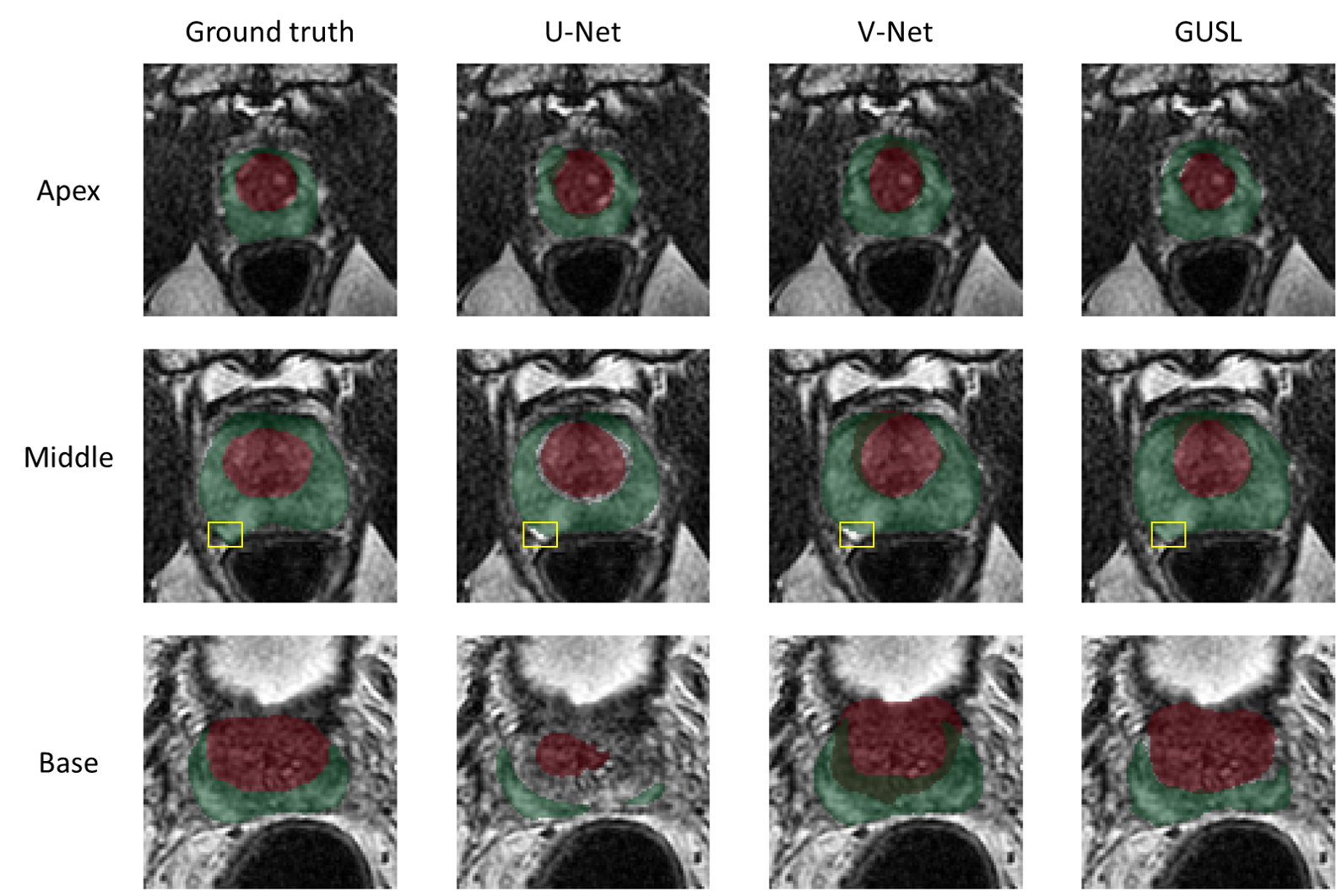}
\caption{Qualitative comparison between GUSL and other DL methods on the zonal segmentation task for T2 Cube MRIs in the USC-Keck dataset. PZ and TZ segmentations are in green and red, respectively.} \label{fig:keck_t2c}
\end{center}
\end{figure*}
%%%%%%%%%%%%%%%%%%%%%%%%

In Figure \ref{fig:keck_t2c}, we provide qualitative comparisons for the zonal segmentation task, dividing into the same regions as for the gland segmentation. The red mask illustrates the TZ segmentation, while the green mask the PZ. 

For the apex of the prostate, U-Net failed to predict the upper PZ region, while both V-Net and GUSL performed better, with GUSL being able to segment more accurately the lower parts of the PZ. Moreover, in the middle section, the U-Net predictions have under-segmented both the PZ and TZ areas, compared to V-Net and GUSL. Also, one can notice that GUSL achieves a better result in delineating the PZ zone along its bottom boundaries with the rectum as shown in the yellow box. Also, V-Net has more overlapping areas predicted as PZ and TZ, compared to GUSL. Also, as in the gland segmentation task, the performance in the prostate base is poorer for all models. U-Net has given a very under-segmented result, missing large parts of the two zones. V-Net's predictions have a high overlap between the two zones, failing to identify correctly their boundaries. However, as we can see, GUSL prediction is more accurate and more aligned with the ground truth, despite some under-segmented areas.

\subsection{Model size comparison}
In Table~\ref{tab:complexity_benchmark}, we include the model size and FLOPS of each method to highlight the green advantages of the proposed methodology when it comes to complexity and model size. GUSL offers a significantly smaller model size, several times less than the DL models under comparison. Also, the complexity is much smaller than most of the models. For example, the 2D U-Net requires almost as twice the number of FLOPS as GUSL, but it achieves a much poorer performance as we saw in the quantitative and qualitative comparisons. These comparisons stress the huge advantages of the GL-based solutions for application deployment and reduce the requirement of expensive hardware.

%%%%%%%%%%%%%%%%%%%%%%%%
\begin{table}[t]
\centering
\scriptsize
\caption{Comparison of model size and complexity in inference between GUSL and the several deep learning baseline models.}
\label{tab:complexity_benchmark}
\begin{tabular}{lcc}
\toprule
       & \# of parameters & FLOPS \\ 
\midrule
V-Net~\citep{milletari2016v}  & 45,603,934 ($\times 40$) & 379B ($\times 53$) \\ 
U-Net++~\citep{zhou2019unet++} & 42,716,150 ($\times 38$) & 1150.3B ($\times 162$)\\ 
3D APA-Net~\citep{jia20193d} & 41,682,078 ($\times 37$) & - \\ 
nnU-Net~\citep{isensee2018nnu} & 31,195,594 ($\times 28$) & 346B ($\times 49$) \\ 
3D GCN~\citep{peng2017large} & 23,587,180 ($\times 21$) & - \\ 
2D U-Net~\citep{unet} & 17,970,626 ($\times 16$) & 13.6B ($\times 2$)\\ 
3D Attention U-Net~\citep{schlemper2019attention} & 5,888,004 ($\times 5$) & 72.8B ($\times 10$)\\ 
\midrule
\textbf{GUSL} & \textbf{1,132,176 ($\times 1$)} & \textbf{7.11B ($\times 1$)}\\ 
\bottomrule
\end{tabular}
\end{table}
%%%%%%%%%%%%%%%%%%%%%%%%

\section{Conclusion}\label{sec:conclusion}

This work proposes a novel and efficient machine learning architecture, named GUSL, for 3D medical image segmentation, applied to prostate gland and zonal segmentation. It is a multi-scale processing model that uses regression in a bottom-up way to initially predict the coarse prostate segmentation, correct its boundaries,and up-sample it to the original resolution. It is a feed-forward model that learns the feature representations using a linear transform. Hence, it is an inherently transparent and interpretable model, which can pave the way for future studies to try to interpret the decision-making process or discover new imaging biomarkers. Extensive experiments in three datasets showed the state-of-the-art performance of our proposed methodology. Moreover, GUSL has the smallest model size and complexity among all the other models in comparison. It is worth noting that our proposed GUSL model provides a generic model for object segmentation, with particular benefits for medical image segmentation. In our future research, we will apply GUSL in different organs and medical images to show that the proposed methodology can have a higher impact within medical imaging, thus offering a more transparent framework for AI-based applications in healthcare.

\printcredits

% \bibliographystyle{model2-names}
% \bibliographystyle{model1-num}
\bibliographystyle{unsrt}

\bibliography{cas-refs}


% Author biography without author photo.
% Author biography. Author biography. Author biography.
% Author biography. Author biography. Author biography.

\end{document}